\documentclass[nofootinbib,twocolumn,nosuperscriptaddress,preprintnumbers,letterpaper,natbib,nofootinbib]{revtex4-2}
\pdfoutput = 1
\usepackage{amssymb}
\usepackage{amsmath}
\usepackage{epsfig}
\usepackage[usenames,dvipsnames]{xcolor}
\usepackage{hyperref}
\usepackage{comment}
\usepackage{cleveref}
\usepackage{multirow}
\usepackage{capt-of}
\usepackage{siunitx}
\usepackage{graphicx}
\usepackage{slashed}
\usepackage{tabularx}
\usepackage{enumitem}
\usepackage{multirow}
\usepackage{braket}
\usepackage{verbatim}
\usepackage{amsfonts}
\usepackage[compat=1.1.0]{tikz-feynman}
\tikzfeynmanset{warn luatex=false}
\usepackage{scalerel}
\usepackage[toc,page]{appendix}
\usepackage[skins,theorems,many]{tcolorbox}
\usepackage{subfigure}
\usepackage{mathrsfs}
\tcbset{highlight math style={enhanced,
  colframe=gray,colback=white,arc=2pt,boxrule=1pt}}

\hypersetup{
    colorlinks=true,       
    linkcolor=Cyan,          
    citecolor=Magenta}

\newcommand{\nocontentsline}[3]{}
\newcommand{\tocless}[2]{\bgroup\let\addcontentsline=\nocontentsline#1{#2}\egroup}
\newcommand{\be}{\begin{equation}}
\newcommand{\ee}{\end{equation}}
\newcommand{\ba}{\begin{array}}
\newcommand{\ea}{\end{array}}
\newcommand{\bea}{\begin{eqnarray}}
\newcommand{\eea}{\end{eqnarray}}
\newcommand{\nn}{\nonumber}

\def\sfrac#1#2{{\textstyle{#1\over #2}}}
\makeatletter

\begin{document}

\title{
The Flavor of QCD Axion Dark Matter
}

\author{Gonzalo Alonso-{\'A}lvarez}
\email{galonso@physics.mcgill.ca}
\thanks{ORCID: \href{https://orcid.org/0000-0002-5206-1177}{0000-0002-5206-1177}}
\author{James M.\ Cline}
\email{jcline@physics.mcgill.ca}
\thanks{ORCID: \href{https://orcid.org/0000-0001-7437-4193}{0000-0001-7437-4193}}
\author{Tianzhuo Xiao}
\email{tianzhuo.xiao@mail.mcgill.ca}
\affiliation{McGill University Department of Physics \& Trottier Space Institute, 3600 Rue University, Montr\'eal, QC, H3A 2T8, Canada}

\begin{abstract} 
\noindent We argue that demanding a consistent cosmological history, including the absence of domain walls and strongly interacting relics at the Peccei-Quinn scale, singles out two concrete realizations of hadronic QCD axions as viable dark matter models.
These realizations generally feature flavor-violating axion couplings to Standard Model quarks that are unsuppressed at low energies.
As a consequence, experiments looking for flavor-violating hadronic processes involving the axion can be sensitive probes of QCD axion dark matter models.
In particular, we show that the NA62 and KOTO experiments could detect the  $K\rightarrow\pi + a$ decay for axions consistent with the observed dark matter abundance via the post-inflationary misalignment mechanism.

\end{abstract}

\maketitle

{
  \hypersetup{linkcolor=black}
 \setlength\parskip{0.0pt}
  \setlength\parindent{0.0pt}
\tableofcontents
}

\section{Introduction}

The QCD axion as a dark matter candidate has received renewed attention in the past few years.  Part of its appeal is that it was invented not for the purpose of providing the dark matter of the universe, but rather to solve the strong CP problem by effectively
promoting the $\theta_{\rm QCD}$ parameter to a dynamical field, that naturally relaxes to zero \cite{Peccei:1977hh,Peccei:1977ur,Weinberg:1977ma,Wilczek:1977pj}.  
Serendipitously, it provides rich possibilities for being produced with the right abundance to be a good cold dark matter candidate, despite its small sub-eV mass.  
Most minimally, it can be produced through the misalignment mechanism, as the axion field at the onset of its cosmological evolution is not generally located at the minimum of its potential~\cite{Abbott:1982af,Preskill:1982cy,Dine:1982ah}.
If cosmological inflation occurs at a high scale above that of Peccei-Quinn symmetry breaking, this scenario is completely predictive and singles out a preferred value for the QCD axion mass, thereby fixing the only continuous free parameter of the theory.

The axion is also an appealing dark matter candidate because of its prospects for direct detection, since a certain strength of couplings to standard model particles is generically predicted.  
Exploiting this, an ever-growing program of experimental searches~\cite{Irastorza:2018dyq}, astrophysical observations~\cite{Raffelt:2006cw}, and cosmological probes~\cite{Marsh:2015xka} is underway to look for this dark matter candidate.
However, the details of the interactions of the axion with the different SM species and their relative strength are model dependent~\cite{DiLuzio:2020wdo}, which results in a large parameter space that needs to be thoroughly tested.

The first and most minimal QCD axion model was quickly ruled out because of its strong couplings to light quarks,
leading to unacceptably large neutral current processes like $J/\Psi \to a \gamma$ \cite{Weinberg:1977ma,Wilczek:1977pj}.
This led to the development of ``invisible'' axion models, in which the couplings to light quarks were naturally suppressed.
There exist two main variants: DFSZ models~\cite{Dine:1981rt,Zhitnitsky:1980tq},
where the axion couples to quarks and electrons; and KSVZ models~\cite{JEKim-axion,Shifman:1979if},
in which the only SM states to which the axion couples directly are gluons.  

The KSVZ and DFSZ models are the main benchmarks for QCD axion dark matter searches.
In both scenarios, the standard lore is that axions couple to hadrons in a flavor-conserving way, via the interaction with their constituent quarks and/or through the gluon coupling.
In this work, we argue that the existence of flavor-changing axion couplings to quarks is well motivated by cosmological and model-building considerations.
Given that flavor-violating observables can test axion-hadron couplings more than three orders of magnitude smaller than flavor-conserving ones~\cite{MartinCamalich:2020dfe}, their presence offers excellent prospects to test the QCD axion dark matter paradigm.
In fact, as we will see, terrestrial high-energy physics experiments like NA62~\cite{NA62:2017rwk} and KOTO~\cite{Yamanaka:2012yma} have sensitivity to QCD axions with masses as small as $\sim 100\,\mu\mathrm{eV}$, a range compatible with current dark matter abundance predictions in the post-inflationary misalignment mechanism~\cite{Buschmann:2021sdq,Gorghetto:2020qws,Klaer:2017ond,Kawasaki:2014sqa}.

We start in Section \ref{sect:cosmo} by reviewing
the cosmology of QCD axions, with emphasis on models that naturally avoid the domain wall problem, and that of unwanted relics involving the heavy exotic quarks of
the KSVZ models.  In Section \ref{sect:mixing} we derive the new couplings of axions to light quarks, induced by the
mass mixing with heavy quarks, that is required for 
depleting the forbidden relics.  Section \ref{sect:pheno} derives constraints on the effective
axion-quark couplings from various flavor-changing 
rare decays, and effects of flavor-conserving interactions on stellar evolution.  In Section \ref{sect:concl} we summarize and give conclusions.
Details of mixing angle computations and their effect on unitarity of the CKM matrix are given in the appendices.

\section{QCD axion dark matter cosmology}
\label{sect:cosmo}

There exist many different ways to realize the Peccei-Quinn (PQ) mechanism, which give rise to models with different particle content and QCD axions with different properties.
However, if the QCD axion is to make up the observed dark matter of the universe, the strict requirement of a consistent cosmological history singles out a few preferred models.

\subsection{Domain wall number}

Although other alternatives exist~\cite{Turner:1985si,Choi:1996fs,Alonso-Alvarez:2017hsz,Nelson:2018via,Graham:2018jyp,Takahashi:2018tdu,Alonso-Alvarez:2019ixv,Co:2020xlh,DiLuzio:2021gos}, the simplest and most predictive scenario for the production of QCD axion dark matter is the post-inflationary misalignment mechanism~\cite{Preskill:1982cy}.
In this paradigm, the $U(1)$ PQ global symmetry is spontaneously broken at the critical temperature $T_{\rm PQ}\sim f_{\rm PQ}$, which is assumed to be lower than the reheating temperature of the universe.
As a consequence, the complex PQ field acquires a vacuum expectation value $\braket{|\Phi|}\sim f_{\rm PQ}$ in the radial direction, and a phase that is uncorrelated in different causal patches of the universe.
That is, the value of the angular component of the PQ field, the axion, is randomized on scales larger than $H^{-1}(T_{\rm PQ})$.

As the universe expands, different disconnected patches enter the causal horizon giving rise to topologically protected configurations where the $U(1)$ field winds around the origin of its potential.
These are global cosmic strings, which are stable as long as the PQ symmetry remains unbroken and the angular axion degree of freedom does not have a potential.
At a temperature around the QCD phase transition, $T_{\rm QCD}\sim \Lambda_{\rm QCD}$, nonperturbative QCD effects generate a potential for the axion
\begin{equation} \label{eq:axion_potential}
    V_a \simeq \Lambda_{\rm QCD}^4 \left[ 1 - \cos\left(N_{\rm DW}\frac{a}{f_{\rm PQ}} \right) \right]\,,
\end{equation}
and explicitly break the $U(1)$ symmetry into a discrete $Z_{N_{\rm DW}}$ group.
If $N_{\rm DW}>1$, 
this symmetry breaking generates $N_{\rm DW}$ inequivalent, degenerate vacua, and thus leads to the formation of domain walls between different regions of space populating each of them.
$N_{\rm DW}$ is referred to as the domain wall number of the model.

Hence, models with $N_{\rm DW} > 1$ lead to stable domain wall networks that are cosmologically unacceptable, as their redshift scaling quickly leads them to dominate the energy density of the universe.
This rules out such models as viable QCD axion dark matter scenarios.
Modifications that can make the domain walls unstable can potentially rescue $N_{\rm DW} > 1$, but they are nontrivial to successfully implement.
The simplest possible solutions of introducing a tilt in the axion potential that breaks the degeneracy between the vacua~\cite{Sikivie:1982qv,Larsson:1996sp}, or a bias in the initial axion field distribution~\cite{Larsson:1996sp,Hindmarsh:1996xv,Coulson:1995nv}, have been shown to be problematic~\cite{Ringwald:2015dsf,Gonzalez:2022mcx,Beyer:2022ywc}. Although other possibilities have been proposed~\cite{Lazarides:1982tw,Stojkovic:2005zh,Kawasaki:2015lpf,Sato:2018nqy,Ferrer:2018uiu,Caputo:2019wsd}, the constructions are far from minimal.

The potential in Eq.~\eqref{eq:axion_potential} in models with $N_{\rm DW} = 1$ 
feature a single vacuum.
That said and due to the presence of cosmic strings at the time when the axion potential arises around the QCD phase transition, domain walls form in this scenario too.
However, the $N_{\rm DW} = 1$ case is radically different because this string-domain wall network is unstable and decays quickly after the axion potential is generated~\cite{Chang:1998tb}.
This case is therefore free of the aforementioned cosmological problem.
We conclude that only QCD axion models with $N_{\rm DW}=1$ are viable dark matter candidates, unless extensive modifications to the axion theory or the cosmological history are postulated.
In particular, this disfavors DFSZ-type models,\footnote{One can couple the axion to a single generation of quarks to ensure $N_{\rm DW}=1$, but in that case flavor-violating axion couplings to quarks arise~\cite{Peccei:1986pn,Krauss:1986wx,DiLuzio:2017ogq}. We leave the study of post-inflationary axion DM possibilities in that scenario for future work.} which have $N_{DW}=6$, and restricts KSVZ-type models to contain a single exotic quark carrying PQ charge.

\subsection{Dark matter density}

QCD axion models have a single continuous free parameter, the scale of PQ symmetry breaking $f_{\rm PQ}$.
It is customary to define the axion decay constant as $f_a = f_{\rm PQ}/N_{\rm DW}$, so for models without a domain wall problem $f_a = f_{\rm PQ}$ and we can use them interchangeably.
The mass of the axion is determined by $f_a$ and can be extracted from the potential in Eq.~\eqref{eq:axion_potential}, leading to~\cite{GrillidiCortona:2015jxo}
\begin{equation}
    m_a = 5.70\times10^{-5}\,\mathrm{eV} \, \left( \frac{10^{11}\,\mathrm{GeV}}{f_a} \right).
\end{equation}
The axion decay constant is thus the single parameter that determines the energy density in axions produced through the misalignment mechanism.
Although calculating the contribution from the axion zero mode (or average initial misalignment) is straightforward, the decay of the axion string-wall network releases a large amount of energy in the form of mildly relativistic axions that must also be accounted for.
In fact, recent calculations agree that this contribution to the axion relic abundance dominates over the one coming from the zero mode~\cite{Kawasaki:2014sqa,Klaer:2017ond,Gorghetto:2020qws,Buschmann:2021sdq}.

In order to determine the spectrum of axions produced from the decay of the string-wall network, one must numerically track the evolution of the cosmic strings from their formation at $T_{\rm PQ}\sim f_{a}$ to their decay at $T_{\rm QCD}\sim \Lambda_{\rm QCD}$.
This is an extremely challenging problem due to the large hierarchy of scales between the size of the strings $\sim H^{-1}(T_{\rm PQ})$ and their typical relative separation $\sim H^{-1}(T_{\rm QCD})$. 
The calculation has been done by a number of groups using various numerical approximations and extrapolations, leading to results in the $m_a = 10^{-5}-10^{-3}\,\mathrm{eV}$ range, with significant dispersion~\cite{Kawasaki:2014sqa,Klaer:2017ond,Gorghetto:2020qws,Buschmann:2021sdq}.
The preferred range of axion masses and corresponding decay constants is summarized in Table~\ref{tab:axion_density}.

Thus, the requirement of matching the observed dark matter abundance further restricts the QCD axion parameter space.
Current calculational efforts point to models with a decay constant in the  $f_a = 5\times10^{9}-3\times10^{11}\,\mathrm{GeV}$ range.

\begin{table}[t]\centering
\setlength\tabcolsep{5pt}
\def\arraystretch{1.5}
\begin{tabular}{ c  c  c }\hline\hline
 Ref. & Mass $[\mu\mathrm{eV}]$ & Decay constant $[10^{10}\,\mathrm{GeV}]$ \\
\hline
\cite{Buschmann:2021sdq} & $40 < m_a < 180$ & $3.2 < f_a < 14$ \\
\cite{Gorghetto:2020qws} & $500 < m_a$ & $f_a < 1.1$ \\
\cite{Klaer:2017ond} & $22.8 < m_a < 29.9$ & $19.3 < f_a < 25.0$ \\
\cite{Kawasaki:2014sqa} & $80 < m_a < 130$ & $4.4 < f_a < 7.1$ \\
\hline\hline
\end{tabular}
\caption{Predictions for the QCD axion dark matter mass and decay constant from numerical simulations of string networks.
All the results are for $N_{\rm DW}=1$ and order chronologically from most to least recent.
The bottom two results include the (subdominant) domain wall contribution while the upper two simulations are strings-only.}
\label{tab:axion_density}
\end{table}

\subsection{Exotic strongly interacting relics}
In order for the PQ global symmetry to be anomalous under QCD and thus solve the strong CP problem, there must exist colored fermions chirally charged under $U(1)_{\rm PQ}$, that transmit the anomaly.
If such fermions are the SM quarks, one obtains DFSZ-type models.
If the SM quarks are not charged under $U(1)_{\rm PQ}$, it is necessary to introduce additional (heavy) colored states, leading to KSVZ-type or hadronic axion models. 
As mentioned above, avoiding the domain wall problem requires dark matter models to have $N_{DW}=1$, and thus a single PQ-charged strongly interacting fermion.
This rules out DFSZ models, under which all six SM quarks are charged, so in what follows we focus on KSVZ scenarios.

Hadronic axion models with $N_{DW}=1$ feature a single pair of heavy fermions $(Q_L,\,Q_R)$ which are vectorlike under the SM gauge groups but are chiral under $U(1)_{\rm PQ}$.
They must transform in the fundamental representation of $SU(3)_c$ and be a singlet under $SU(2)_L$ to avoid higher multiplicity, and thus the possible different models are classified by their $U(1)_Y$ charges.

In the early universe, these heavy fermions are produced in the thermal plasma through the strong interactions.
Depending on their gauge charges, they may form stable charged or neutral exotic hadrons whose present abundance is severely constrained by observations.
In order to avoid that, their interactions must allow a sufficiently fast decay into SM states.
The detailed analysis of Ref.~\cite{DiLuzio:2017pfr} concludes that the lifetimes $\lesssim 0.01\,\mathrm{s}$ are needed for the heavy fermion decays to be compatible with BBN and other cosmological processes.
This can only be achieved if their $U(1)_Y$ charges allows them to couple to SM states via operators of dimension $\leq 5$, yielding
another selection criterion that further shrinks the landscape of possible QCD axion dark matter models.

\subsection{Summary: viable models}

To summarize, we have argued that QCD axion models must satisfy three requirements if the axion is to constitute the dark matter:
\begin{enumerate}
\item They need $N_{\rm DW}=1$ to avoid a domain wall problem.
\item Any new strongly interacting fermion requires gauge charges allowing it to couple to SM states with operators of dimension $\leq 5$.
\item The axion decay constant must be in the range $f_a = 5\times10^{9}-3\times10^{11}\,\mathrm{GeV}$ to reproduce the observed dark matter abundance via misalignment.
\end{enumerate}
It is easy to see that these three conditions can only be simultaneously satisfied in hadronic axion models with a single additional heavy fermion in either of the SM gauge group representations 
\begin{align}
\label{R1R2def}
\mathrm{KSVZ{\rm-}I} &: (3,\,1,\,-1/3),\,\mathrm{or} \nn \\
\mathrm{KSVZ{\rm-}II} &: (3,\,1,\,+2/3).
\end{align}
The heavy vectorlike quarks are thus required to have the same quantum numbers as the SM down-type or up-type right-handed quarks.
Since they are chirally charged under $U(1)_{\rm PQ}$, their mass must arise from their interaction with the complex PQ field.
By normalizing the PQ charge of $\Phi$ to be 1, we can take\footnote{One could more generally consider $\Delta \chi = k$ and have the mass of $Q$ arise from a higher dimensional operator involving $\Phi^k$, but such complication does not lead to any qualitative difference.} $\Delta \chi = \chi_L - \chi_R = 1$, where $\chi_{L/R}$ is the PQ charge of $Q_{L/R}$, and thus write
\begin{equation}\label{eq:Q_mass_term}
\mathcal{L} \supset y_{Q} \Phi \bar{Q}_L Q_R + \mathrm{H.c.}.
\end{equation}
After PQ symmetry breaking, this term produces a mass for the heavy quarks $m_Q = y_Q f_a$.
As we now argue, this is a predictive scenario that can lead to novel phenomenology due to the interactions between the axion, the heavy $Q$, and the SM quarks.


\section{Flavor-violating QCD axion dark matter}
\label{sect:mixing}

The discussion in the previous section has shown that QCD axion dark matter models necessarily introduce a heavy vector-like quark with SM right-handed quark quantum numbers.
Having the same quantum numbers, mass mixing between the SM and the heavy quarks is possible.
As we argue next, mass mixing is in fact unavoidably induced by the operators necessary for $Q$ decay.

Let us consider the example of KSVZ-I; the case of KSVZ-II is completely analogous under the exchanges $u_R\leftrightarrow d_R$, $H\leftrightarrow \Tilde{H}$.
Depending on the PQ charge assignment of the two chiral components of $Q$, four distinct operators of dimension $d\leq 4$ are possible: 
\begin{align} \label{eq:operators}
    \mathcal{O}_4^M &= M_d\bar{Q}_L d_R, &\mathrm{for}& \quad (\chi_L,\,\chi_R) = (0,\,-1), \nonumber\\
    \mathcal{O}_4^H &= y_{1,d}H\bar{d}_L Q_R, &\mathrm{for}& \quad (\chi_L,\,\chi_R) = (1,\,0), \nonumber\\
    \mathcal{O}_4^\Phi &= y_{2,d}\Phi\bar{Q}_L d_R, &\mathrm{for}& \quad (\chi_L,\,\chi_R) = (1,\,0),\nonumber\\
    \mathcal{O}_4^{\Phi^\dagger} &= y_{3,d}\Phi^\dagger\bar{Q}_L d_R, &\mathrm{for}& \quad (\chi_L,\,\chi_R) = (-1,\,-2).
\end{align}
Here, $y_{n,d}$ represents Yukawa couplings of quarks with flavor $d=d,s,b$. 
At dimension $5$, additional operators are possible,
\begin{align} \label{eq:operators_d5}
    \mathcal{O}_5^{|H|} &= \frac{\lambda_d}{\Lambda} |H|^2\bar{Q}_L d_R, &\mathrm{for}& \quad (\chi_L,\,\chi_R) = (0,\,-1), \nonumber\\
    \mathcal{O}_5^{|\Phi|} &= \frac{\lambda'_d}{\Lambda} |\Phi|^2\bar{Q}_L d_R, &\mathrm{for}& \quad (\chi_L,\,\chi_R) = (0,\,-1), \nonumber\\
    \mathcal{O}_5^H &= \frac{\lambda_{1,d}}{\Lambda} \Phi H\bar{d}_L Q_R, &\mathrm{for}& \quad (\chi_L,\,\chi_R) = (0,\,-1), \nonumber\\
    \mathcal{O}_5^\Phi &= \frac{\lambda_{2,d}}{\Lambda} \Phi^2 \bar{Q}_L d_R, &\mathrm{for}& \quad (\chi_L,\,\chi_R) = (2,\,1),\nonumber\\
    \mathcal{O}_5^{\Phi^\dagger} &= \frac{\lambda_{3,d}}{\Lambda}\left(\Phi^\dagger\right)^2\bar{Q}_L d_R, &\mathrm{for}& \quad (\chi_L,\,\chi_R) = (-2,\,-3),
\end{align}
where $\lambda_{n,d}$ are dimensionless couplings and $\Lambda$ is a dimensionful scale.
However, after symmetry breaking, their phenomenology is equivalent to the dimension-4 operators upon substituting $\lambda v^2/\Lambda\rightarrow M$ and $\lambda' f_a^2/\Lambda\rightarrow M$ for the first two, or $\lambda f_a /\Lambda \rightarrow y$ for the latter three.
The only relevant distinction arises from the different PQ charge assignments, as we will see.
For simplicity, in what follows we use the notation of the dimension-$4$ operators but while noting that our results equally apply to the dimension-$5$ ones with the aforementioned substitutions.

In the absence of a flavor symmetry, operators involving any of the three generations of SM quarks are possible, and thus the Yukawa couplings $y$ and the mass parameter $M$ should be understood as vectors in flavor space.
After the Higgs and PQ scalar fields acquire VEVs, all four operators induce mass mixing between the heavy and the SM quarks.
We will show that this mixing can lead to flavor-violating axion-SM quark couplings.
  
Consider first the case of $\mathcal{O}_4^\Phi$;
the $\mathcal{O}_4^{\Phi^\dagger}$ operator is completely analogous, as is $\mathcal{O}_4^M$, after the replacement $y_{2,d}\rightarrow M_d$.
For simplicity, we assume the $y_{2,d}$ couplings to be defined in the mass basis of SM quarks, after the unitary rotation associated with the CKM matrix has been performed.
After spontaneous breaking of the PQ symmetry, the following $4\times 4$ mass matrix arises:
\begin{equation}
    \begin{pmatrix}
    \bar{d}_L & \bar{s}_L & \bar{b}_L & \bar{Q}_L
    \end{pmatrix}
    \begin{pmatrix}
    m_d & 0 & 0 & 0 \\
    0 & m_s & 0 & 0 \\
    0 & 0 & m_b & 0 \\
    y_{2,d}f_a & y_{2,s}f_a & y_{2,b}f_a & m_Q
    \end{pmatrix}
    \begin{pmatrix}
    d_R \\
    s_R \\
    b_R \\
    Q_R
    \end{pmatrix},
    \label{mixmat}
\end{equation}
where $m_q=y_q v$, with $v=174$~GeV being the complex Higgs VEV. 
The PQ VEV for $N_{\rm DW}=1$ models is given by $f_a = f_{\rm PQ}$.
Recall that the heavy quarks obtain their masses at the scale of $f_a$, and we can thus write $m_Q = y_Q f_a$ following Eq.~\eqref{eq:Q_mass_term}.
The axion phase in the mass matrix in Eq.~\eqref{mixmat} has been removed by an appropriate (anomalous) chiral rotation of the $(Q_L,\,Q_R)$ field.

Assuming that $m_d\ll m_Q,\,y_{2,d}f_a$, we diagonalize the system perturbatively using unitary matrices $U_L$ and $U_R$ such that $U_L^\dagger M M^\dagger U_L = \Lambda^2 = U_R^\dagger  M^\dagger M U_R$.
The mass eigenvalues are
\begin{align}
    \lambda_Q^2 &= m_Q^2 + \sum_d y_{2,d}^2 f_a^2, \\
    \lambda_d^2 &= m_d^2 \left( 1 - \frac{y_{2,d}^2 f_a^2}{\lambda_Q^2} \right),
\end{align}
to leading order in the $m_d/m_Q$ expansion.
The light quark masses are therefore scaled by a flavor-dependent factor $\leq 1$.
It will be shown that observational constraints require this correction to be small, implying $y_{2,d}f_a / m_Q \ll 1$.
We thus henceforth assume that $m_d \ll y_{2,d}f_a \ll m_Q$ to simplify the analytic expressions.

The mixings of the left- and right-handed fields, respectively, are given by
\begin{align}
    \begin{pmatrix}
    d_{L/R}' \\
    Q_{L/R}'
    \end{pmatrix}
    = \begin{pmatrix}
    \cos\theta_{Qd}^{L/R} & \sin\theta_{Qd}^{L/R} \\
    -\sin\theta_{Qd}^{L/R} & \cos\theta_{Qd}^{L/R}
    \end{pmatrix}
    \begin{pmatrix}
    d_{L/R} \\
   Q_{L/R}
    \end{pmatrix},
\end{align}
with the mixing angles
\begin{align}
    \tan 2\theta_{Qd}^L &\simeq -2 \frac{m_d \,y_{2,d}^* f_a}{m_Q^2}, \\
    \tan 2\theta_{Qd}^R &\simeq 2 \frac{m_Q \, y_{2,d}^* f_a}{m_Q^2 - m_d^2}\,.
\end{align}

\begin{figure}[t]
    \centering
    \includegraphics[width=0.65\linewidth]{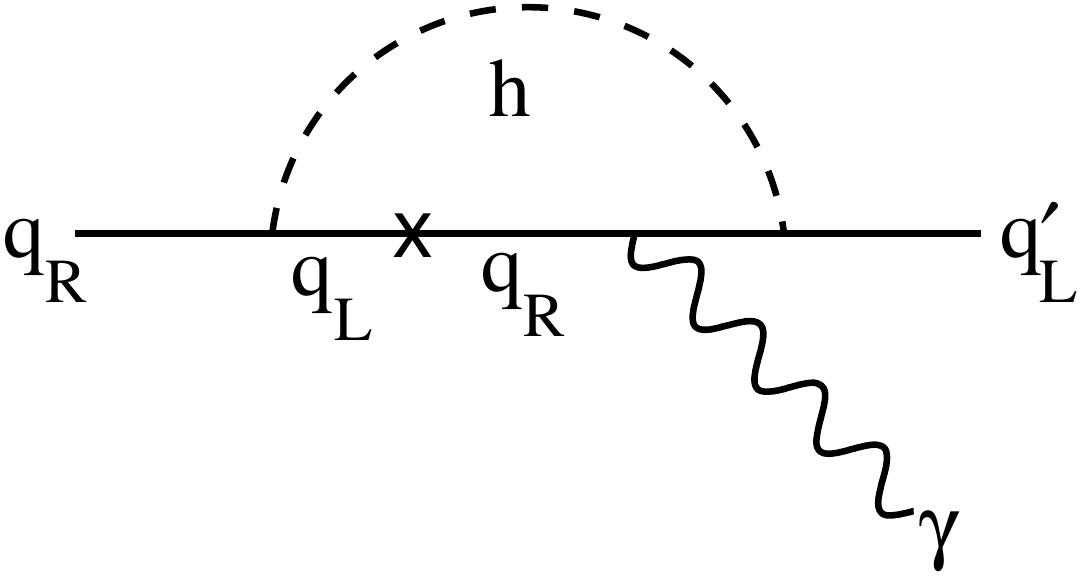}
    \caption{$
    q\to q'\gamma$ contribution from off-diagonal Higgs coupling.}
    \label{fig:bsg}
\end{figure}

Before mixing but after performing the chiral rotation to remove the axion phase from the mass matrix, the axion only interacts with the heavy quark via
\begin{equation}\label{eq:Ca_before_mixing}
    \mathcal{L}\supset - \frac{1}{2f_a} \,\partial_\mu a \, \left( \chi_L \bar{Q}_L \gamma^\mu Q_L +  \chi_R \bar{Q}_R \gamma^\mu Q_R \right).
\end{equation}
Recall that the PQ charge assignments of the left- and right-handed $Q$ components are different for each of the $\mathcal{O}_4^\Phi$, $\mathcal{O}_4^{\Phi^\dagger}$, and $\mathcal{O}_4^M$ operators and their dimension $5$ counterparts. 
After mixing, Eq.~\eqref{eq:Ca_before_mixing} becomes
\begin{align}
    \mathcal{L}\supset - \frac{1}{2f_a} \,\partial_\mu a \, 
    &\left[
    \chi_L \begin{pmatrix}
    \bar{q}_L \, \bar{Q}_L
    \end{pmatrix} 
    C_a^L \gamma^\mu
    \begin{pmatrix}
    q_L \\
    Q_L
    \end{pmatrix}
    \right. \nn \\
    &\left.
    + \chi_R \begin{pmatrix}
    \bar{q}_R \, \bar{Q}_R
    \end{pmatrix} 
    C_a^R \gamma^\mu
    \begin{pmatrix}
    q_R \\
    Q_R
    \end{pmatrix}
    \right].
    \label{Lagrange}
\end{align}
The $4\times 4$ matrices $C_a^{L/R}$ describing the axion couplings are in general nondiagonal, meaning that there exist flavor-violating interactions between the axion and the down-type SM quarks.
The elements of $C_a$ involving SM quarks are
\begin{align}\label{eq:CaL}
    (C_a^{L})_{dd'} &\simeq \frac{ m_d m_{d'} {y_{2,d}^*}\,y_{2,d'}^{\phantom{*}}\,f_a^2}{m_Q^4}, \\
    (C_a^{R})_{dd'} &\simeq \frac{{y_{2,d}^*}\,y_{2,d'}^{\phantom{*}}\,f_a^2}{m_Q^2}.
    \label{eq:CaR}
\end{align}
Perturbative unitarity requires that $f_a / m_Q = 1/y_Q \gtrsim \sqrt{3/(8\pi)}$
\cite{Allwicher:2021rtd}.
Unless $Q$ mixes with just a single quark flavor, the axion has flavor-violating couplings, which
are suppressed by $m_d^2/m_Q^2$ if only the left-handed exotic quark carries PQ charge, but they can be sizable if $\chi_R\neq 0$.
Among the possible PQ charge assignments listed in Eqs.~\eqref{eq:operators} and~\eqref{eq:operators_d5}, only one has $\chi_R=0$.
Thus, for all the other models, low-energy flavor-violating axion couplings to quarks can be significant.

Following a similar procedure for the $\mathcal{O}^H_4$ operator, the induced mass  matrix is
\begin{equation}
    \begin{pmatrix}
    \bar{d}_L & \bar{s}_L & \bar{b}_L & \bar{Q}_L
    \end{pmatrix}
    \begin{pmatrix}
    m_d & 0 & 0 & y_{1,d}\, v \\
    0 & m_s & 0 & y_{1,s}\, v \\
    0 & 0 & m_b & y_{1,b}\, v \\
    0 & 0 & 0 & m_Q
    \end{pmatrix}
    \begin{pmatrix}
    d_R \\
    s_R \\
    b_R \\
    Q_R
    \end{pmatrix}.
\end{equation}
Assuming that $m_d,\,y_{1,d}v \ll m_Q$, the eigenvalues of the squared mass matrix are
\begin{align}
    \lambda_d^2 &= m_d^2\left(1-\frac{y_{1,d}^2 v^2}{m_Q^2}\right), \nonumber\\
    \lambda_Q^2 &= m_Q^2 \left( 1 + \sum_d \frac{y_{1,d}^2 v^2}{m_Q^2} \right),
\end{align}
to leading order in the $m_d/m_Q$ expansion.
The axion couplings to the SM quarks can be extracted as in the previous case, leading to
\begin{align}
    (C_a^L)_{dd'} &= \frac{y_{1,d}^*y_{1,d'}v^2}{m_Q^2},\\
    (C_a^R)_{dd'} &= \frac{m_d m_{d'} y_{1,d}^*y_{1,d'}v^2}{m_Q^4}.
\label{ReEl}
\end{align}
These couplings and the shifts in the SM quark masses are 
further suppressed in this case  by the small ratio $v^2/f_a^2$.
Therefore the $\mathcal{O}^H_4$ and $\mathcal{O}^H_5$ operators do not lead to any significant effects and we neglect them in the following analysis. 

\section{Flavor phenomenology}
\label{sect:pheno}

\begin{figure}[t]
    \centering
    \includegraphics[width=0.8\linewidth]{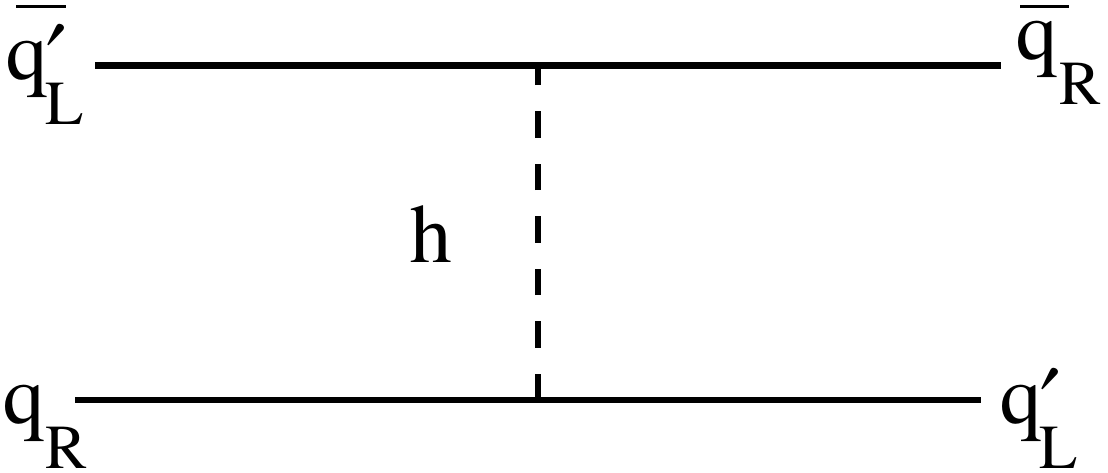}
    \caption{Four-quark  flavor-violating operator induced by Higgs exchange.}
    \label{fig:htree}
\end{figure}

As discussed above, the mass mixing between the heavy $Q$ and the SM quarks introduces flavor-violating axion couplings at low energies.
Thus we expect that processes like the decay $K\to \pi a$ may provide strong contraints on the QCD axion.
In addition, the models under consideration induce other flavor-violating processes not involving the axion, and therefore independent of $f_a$, that we study first.

\subsection{Flavor-violating Higgs couplings}
Mixing with the heavy quark $Q$ gives rise to  new contributions to the SM quark mass  matrix that are not associated with the Higgs mechanism.
Thus, unless the heavy quark mixes with a single light mass eigenstate, flavor-violating couplings of the Higgs boson to the SM quarks arise.
For example, in the model of Eq.\ (\ref{mixmat}) (KSVZ-I representation and $\mathcal{O}_4^\Phi$ operator), the  mixing angles for the light left- and right-handed quarks amongst themselves take the form\footnote{The KSVZ-II models induce an analogous mixing between the up-type quarks.} (see appendix~\ref{sec:quark_mixing_angles})
\begin{align}
\label{thetaL}
\theta_{dd'}^L &\simeq \frac{m_d m_{d'}}{m^2_d-m^2_{d'}}\frac{y^*_{2,d} y_{2,d'} f_a^2}{m_Q^2}\,,\\
\theta_{dd'}^R &\simeq {m^2_d\over m^2_d-m^2_{d'}}\frac{y^*_{2,d} y_{2,d'} f_a^2}{m_Q^2}\,.
\end{align}
While the left-handed mixing of light quarks is unitary to leading order in the $m_d/m_Q$ expansion, that of the right-handed quarks is not.
Right-handed quarks can mix significantly with the heavy $Q$ state, and deviations from unitarity are only suppressed by $(y_{2,d}f_a/m_Q)^2$.  
This has no impact on unitarity of the CKM matrix,
which is determined solely by the left-handed mixing.

\begin{table}[t]\centering
\setlength\tabcolsep{5pt}
\def\arraystretch{1.5}
\begin{tabular}{ c  c  c c }\hline\hline
 & $C_2,\tilde{C}_2\,[\mathrm{TeV}]^{-2}$ & $C_4\,[\mathrm{TeV}]^{-2}$ & $(C_a^R)_{qq'}$ \\
\hline
$\mathrm{Re}\,C_K \, (sd)$ & $7.9\times 10^{-9}$ & $1.6\times 10^{-9}$ & $4.1\times10^{-2}$ \\
$\mathrm{Im}\,C_K \, (sd)$ & $2.1\times 10^{-11}$ & $4.5\times 10^{-12}$ & $2.1\times 10^{-3}$ \\
$|C_{B_d}| \, (bd)$ & $2.8\times 10^{-7}$ & $8.8\times 10^{-8}$ &  $5.5\times 10^{-3}$\\
$|C_{B_s}| \, (bs)$ & $3.1\times 10^{-6}$ & $1.0\times 10^{-6}$ & $1.8\times 10^{-2}$ \\
$\mathrm{Im}\,C_D \, (cu)$ & $1.7\times 10^{-9}$ & $4.5\times 10^{-10}$ & $1.4\times 10^{-3}$ \\
\hline
$t\rightarrow h j \, (tu)$ & --- & --- & $9.5\times 10^{-2}$ \\
$t\rightarrow h j \, (tc)$ & --- & --- & $9.1\times 10^{-2}$ \\
\hline\hline
\end{tabular}
\caption{95\% CL on the coefficients $(C_a^R)_{qq'}$ from flavor-violating quark-Higgs couplings.
The first five rows show the limits on the Wilson coefficients of the effective Hamiltonian Eq.~\eqref{eq:eff_H}, as calculated in~\cite{Bona:2022zhn}, with the derived bound on the flavor-violating axion coupling in the rightmost column.
The bottom lines show bounds on the top quark couplings arising from $t\rightarrow h u$ and $t\rightarrow h c$ decays constrained by the ATLAS search~\cite{ATLAS:2018jqi}.}
\label{tab:UTfit}
\end{table}

The new contributions to flavor mixing generate off-diagonal Higgs couplings
\begin{align}
    y \, H\, \bar d_L \, d_R + \mathrm{H.c.} &\rightarrow [\theta^{L\dagger} y + y\theta^R]_{dd'}\,H\,\bar d_L d'_R + {\rm H.c.} \nn \\
    &\cong y_d \, (C_a^R)_{dd'} H\, \bar d_L \, d'_R + \mathrm{H.c.},
    \label{higgs-mix}
\end{align}
where $H= h/\sqrt{2}$ is the complex Higgs field and $y$ denotes the SM Yukawa coupling matrix in the mass basis.
In the second line of Eq.~\eqref{higgs-mix} we have
rewritten the Higgs couplings in terms of $(C_a^R)_{dd'}$ using Eq.~\eqref{eq:CaR}.
This expression is valid as long as phases can be ignored.
Flavor changing neutral current processes  constrain the $(C_a^R)_{dd'}$ coefficients, which encode the flavor-violating couplings of the axion.

\begin{figure*}[t]
\centering
\centerline{\includegraphics[width=0.5\linewidth]{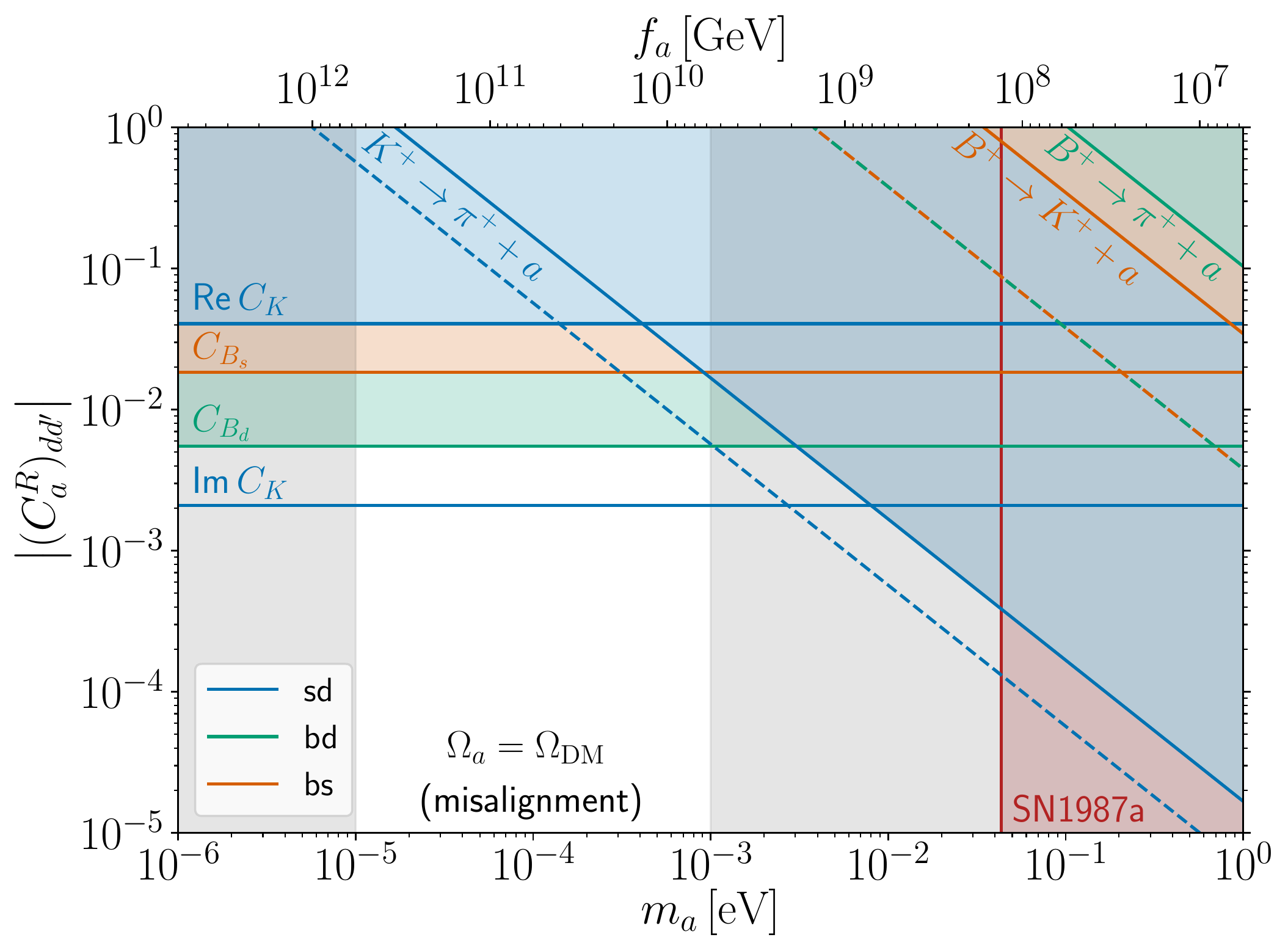}
\includegraphics[width=0.5\linewidth]{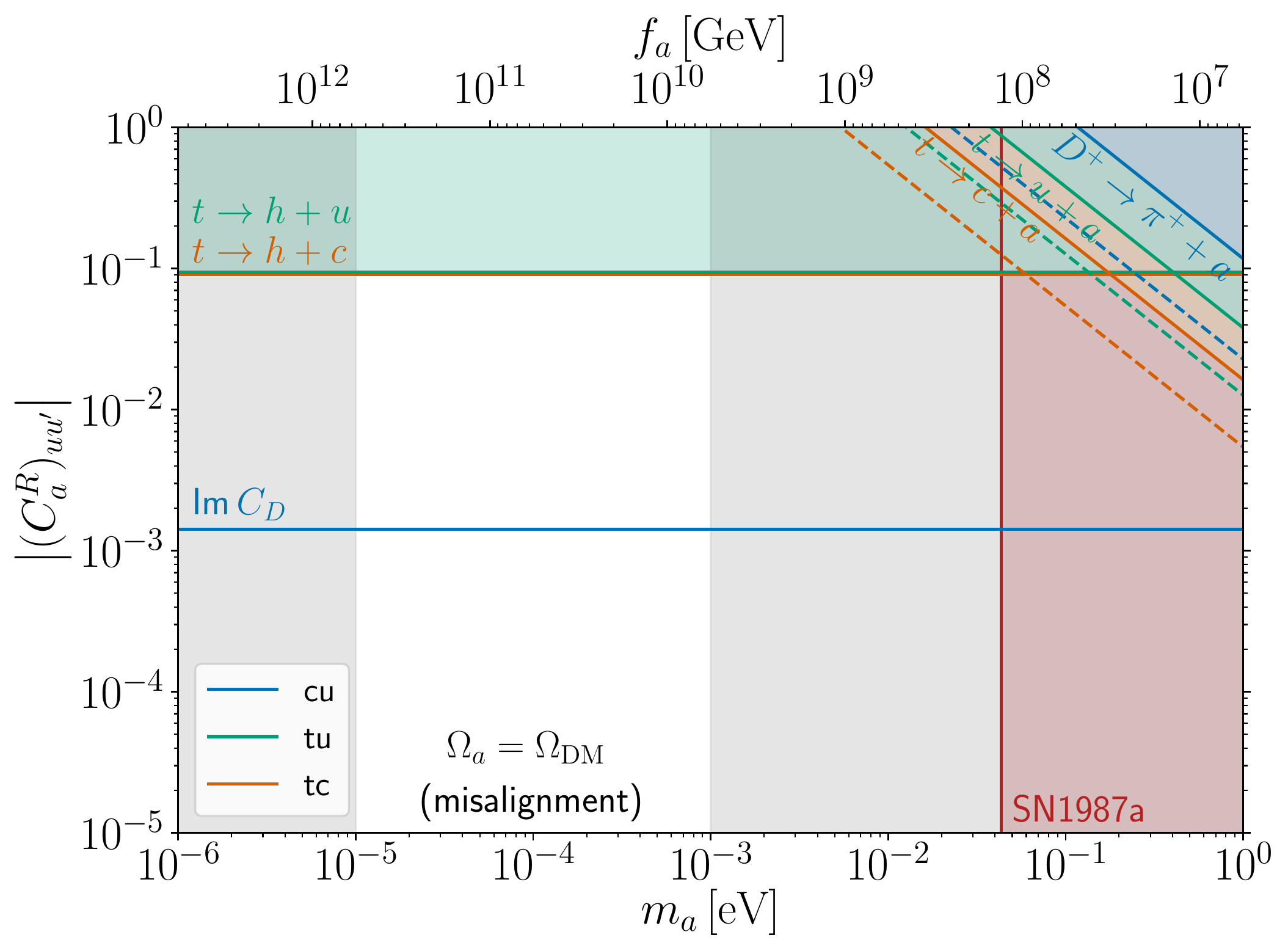}}
\caption{Existing limits and projected sensitivities of searches in the $m_a$ (or $f_a$) vs $C_a^{R}$ parameter space, for down-type quarks (left) and up-type quarks (right).
Different colors represent different flavor combinations.
Horizontal bounds arise from flavor-violating Higgs couplings (this work; see Table~\ref{tab:UTfit}), and oblique ones correspond to flavor-violating meson decays involving the axion (dashed lines show future projections) \cite{MartinCamalich:2020dfe}.
The region where the QCD axion can make up the observed dark matter abundance~\cite{Buschmann:2021sdq,Gorghetto:2020qws,Klaer:2017ond,Kawasaki:2014sqa} is  white (see Table~\ref{tab:axion_density}), while the red shading shows the limit from SN1987a on the axion-gluon coupling \cite{Carenza:2019pxu}.}

\label{fig:Ca_fa}
\end{figure*}

First, these couplings can induce radiative $q \rightarrow q' + \gamma$ quark decays via the one-loop process shown in Fig.\ \ref{fig:bsg}.
Considering the $b\rightarrow s + \gamma$ transition, for which the partonic calculation is a good approximation, the transition magnetic moment is of order
\be
    \mu_{bs} \sim {e y_b^2(C_a^R)_{bs} m_b\over 32\pi^2 m_h^2},
\ee
where we have neglected the the contribution proportional to $y_s \ll y_b$.
To obtain a limit, we match this to the corresponding operator in the expansion of Ref.\ \cite{Descotes-Genon:2015uva}, finding a Wilson coefficient
\be
    \delta C_7 = {y_{b}^2 (C_a^R)_{bs}\over
        4\sqrt{2}\, G_F m_h^2 V_{tb} V^*_{ts}} \cong 0.014\,(C_a^R)_{bs}\,.
\ee
New physics contributions are bounded by $|\delta C_7|\lesssim 0.04$ at $95\%$ CL~\cite{Descotes-Genon:2015uva}, hence this gives the constraint $(C_a^R)_{bs}\lesssim
3$. 
Since this is significantly weaker than the
limit from $B_s$ mixing (see below), we do not attempt to refine the estimate here.

For the analogous transition $s\rightarrow d \gamma$,
the decay $K \rightarrow \pi + \gamma$ does not arise
since the hadronic matrix element of the dipole operator
vanishes for on-shell photons.  Instead one can consider
the decay $K_S \to \pi^0 e^+ e^-$. 
Using methodology of Ref.\ \cite{Cline:2015lqp} (section 7.7), this results in the weak constraint $(C_a^R)_{ds} < 60$. 
Similarly one can estimate that Im$(C_a^R)_{ds}\lesssim 20$ from
the CP-violating decay $K_L \to \pi^0 e^+ e^-$.

Off-diagonal Higgs couplings also induce flavor-changing neutral currents at tree level via the diagram shown in Fig.~\ref{fig:htree}.
As shown in Ref.~\cite{Harnik:2012pb}, integrating out the Higgs leads to the effective Hamiltonian
\begin{equation}\label{eq:eff_H}
H_{\rm eff} = C_2^{qq'} (\bar{q}'_R q_L)^2 + \tilde{C}_2^{qq'} (\bar{q}'_L q_R)^2 + C_4^{qq'} (\bar{q}'_L q_R)(\bar{q}'_R q_L)\,,
\end{equation}
which can be applied to $K^0$, $B^0$, $B^0_s$, and $D^0$ meson mixing.
The full operator basis is defined in Ref.~\cite{UTfit:2007eik}.
The Wilson coefficients in our model can be expressed as
\begin{align}
C_2^{qq'} &= -\frac{(y_q^*)^2}{4 m_h^2} (C_a^R)^2_{q'q}\,, \\
\tilde{C}_2^{qq'} &= -\frac{y_{q'}^2}{4m_h^2} (C_a^R)^2_{q'q}\,, \\
C_4^{qq'} &= -\frac{y_q^*y_{q'}}{2m_h^2} (C_a^R)^2_{q'q}\,,
\end{align}
in terms of the flavor-violating right-handed axion couplings in Eq.~\eqref{eq:CaR}.
The combined fit to all relevant experimental data performed in Ref.\ \cite{UTfit:2007eik} (see latest update in Ref.\ \cite{Bona:2022zhn}) can be used to extract limits on the model parameters, and directly on the $C_a^R$ coefficients.
The 95\% CL limits on the relevant Wilson coefficients are shown in Table~\ref{tab:UTfit}.

The strongest limits on flavor-violating Higgs couplings involving the top quark come from their loop contirbution to the neutron electric dipole moment~\cite{Gorbahn:2014sha}.
However, in our model they apply to the product $|\mathrm{Im}(y_t\, y_u\, (C_a^R)_{tu}(C_a^R)_{ut})|$ which vanishes given that $(C_a^R)_{tu}=(C_a^R)_{ut}^*$; see Eq.~\eqref{eq:CaR}.
In their absence, the leading bounds come from $t\rightarrow q h$ top decays~\cite{Craig:2012vj}.
The rate is given by~\cite{Harnik:2012pb}
\begin{equation}
    \Gamma(t\rightarrow hq) = \frac{y_t^2 |(C_a^R)_{tq}|^2}{64\pi} \frac{(m_t^2 - m_h^2)^2}{m_t^3}\,,
\end{equation}
neglecting the light ($q=c,u$) quark masses.
Using the most recent experimental constraints $\mathscr{B}(t\rightarrow h u)< 1.2\times 10^{-3}$ and $\mathscr{B}(t\rightarrow h c)< 1.1\times 10^{-3}$~\cite{ATLAS:2018jqi} at $95\%$ CL, we obtain the limits shown in the lower two columns of Table~\ref{tab:UTfit}.

\subsection{Flavor-violating axion couplings}

The interactions between the axion and the SM fermions can be written in full generality as
\begin{equation}\label{eq:flavor_couplings}
    \mathcal{L} \supset \frac{\partial_{\mu}a}{2f_a}\bar{f_i}\gamma^{\mu}(c^V_{f_if_j}+c^A_{f_if_j}\gamma_5)f_j\,,
\end{equation}
where $f_{i}$ stands for any lepton or quark species and the axial and vector couplings $c^{A,V}_{f_i f_j}$ are hermitian matrices in flavor space.  
For the QCD axion dark matter models under study, the only sizable couplings are those to right-handed quarks,
\begin{equation}
    c_{qq'}^V = c_{qq'}^A = \frac{1}{2} \chi_R (C_a^R)_{qq'} \,;
\end{equation}
see Eq.~\eqref{eq:CaR}.
The bounds on the different flavor combinations of the couplings in Eq.~\eqref{eq:flavor_couplings} were exhaustively studied in Ref.~\cite{MartinCamalich:2020dfe}, and we base our analysis on those results.

Off-diagonal couplings ($q\neq q'$), can most sensitively be probed using flavor-violating hadron decays.
The limits on $c_{qq'}^V$ arise from two-body charged meson decays and are typically stronger than those on $c_{qq'}^A$.
At present, the most stringent limit out of all the flavor combinations is $c_{sd}^V/f_a \leq 3\times 10^{-12}\,\mathrm{GeV}^{-1}$, coming from charged kaon decays, $K^+\rightarrow\pi^+ + a$, as constrained by the E949 experiment~\cite{E949:2007xyy}.
The first run of the NA62 experiment achieved a marginally weaker sensitivity~\cite{NA62:2020xlg}, and the limit is expected to improve by up to a factor of two in the second run.
In the neutral kaon version of the decay, $K_L\rightarrow\pi^0 + a$, the KOTO experiment is expected to achieve comparable, but somewhat weaker, sensitivity~\cite{Goudzovski:2022vbt}.
Charged $D$ and $B$ meson decays~\cite{CLEO:2008ffk,BaBar:2004xlo,BaBar:2013npw} lead to comparatively weaker bounds on the $bs$, $bd$, and $cu$ flavor combinations.
The $tu$ and $tc$ couplings can only be constrained through their loop-level contribution to the aforementioned decays, leading to bounds with large theoretical uncertainty.
Ref.\ \cite{MartinCamalich:2020dfe} gives a detailed discussion of all the limits summarized above.

Flavor-conserving couplings are best tested through their impact on stellar cooling processes. 
In particular, axion emission from supernovae is sensitive to an effective coupling to nucleons as described in Ref.\ \cite{Carenza:2019pxu}.
The nucleon coupling can be related to the axion-gluon and the different $c_{qq}^A$ flavors by matching to the chiral Lagrangian and taking into account QCD running effects~\cite{GrillidiCortona:2015jxo,DiLuzio:2017ogq,MartinCamalich:2020dfe}.
If all $c_{qq'}^A\lesssim 0.1$, the model-independent gluon contribution alone dominates in the effective nucleon coupling.
In this limit, the duration of the neutrino burst from SN1987a results in a limit $f_a\lesssim1.4\times10^{8}$~GeV~\cite{Carenza:2019pxu}.

The results adapted to the parameter space of the QCD axion models under study are presented in Fig.~\ref{fig:Ca_fa}, in terms of the axion mass (or, equivalently, its decay constant) and the absolute value of the $C_a^R$ coefficients.
The left panel shows the limits on down-type quark couplings, while the right one presents those on up-type ones.
Different colors correspond to different flavor combinations and should be read independently of each other in the absence of any assumption regarding the flavor structure of the operators in Eqs.~\eqref{eq:operators} and~\eqref{eq:operators_d5}.

Being independent of $f_a$, the limits on the flavor-violating Higgs couplings appear as horizontal lines.
For the limits that are sensitive to the phase of $C_a^R$, we show the two extreme cases corresponding to real and complex couplings,  enforcing only the weakest one on $|C_a^R|$.
Conversely, the (flavor conserving) SN1987a limit acts on $f_a$ independently of $C_a^R$ for the values $C_a^R\lesssim 0.1$ of interest.

The rare meson (or top quark) decays constrain the combination $C_a^R / f_a$.
Current bounds are denoted by solid lines and shaded colors, while projections are shown with dashed lines.
Out of all the flavor variations, Fig.~\ref{fig:Ca_fa} demonstrates that only for $sd$ can the bounds on flavor-violating axion couplings access otherwise untested regions of parameter space.
However, for that flavor combination, searches for $K\rightarrow\pi+a$ decay are sensitive tests of the minimal KSVZ-I QCD axion dark matter model.
Saturating the limit on $(C_a^R)_{sd}$, NA62 will probe QCD axion masses as low as $\sim 100\,\mu\mathrm{eV}$, well into the range 
of current dark matter predictions.

\begin{figure}
\centering
\includegraphics[width=\linewidth]{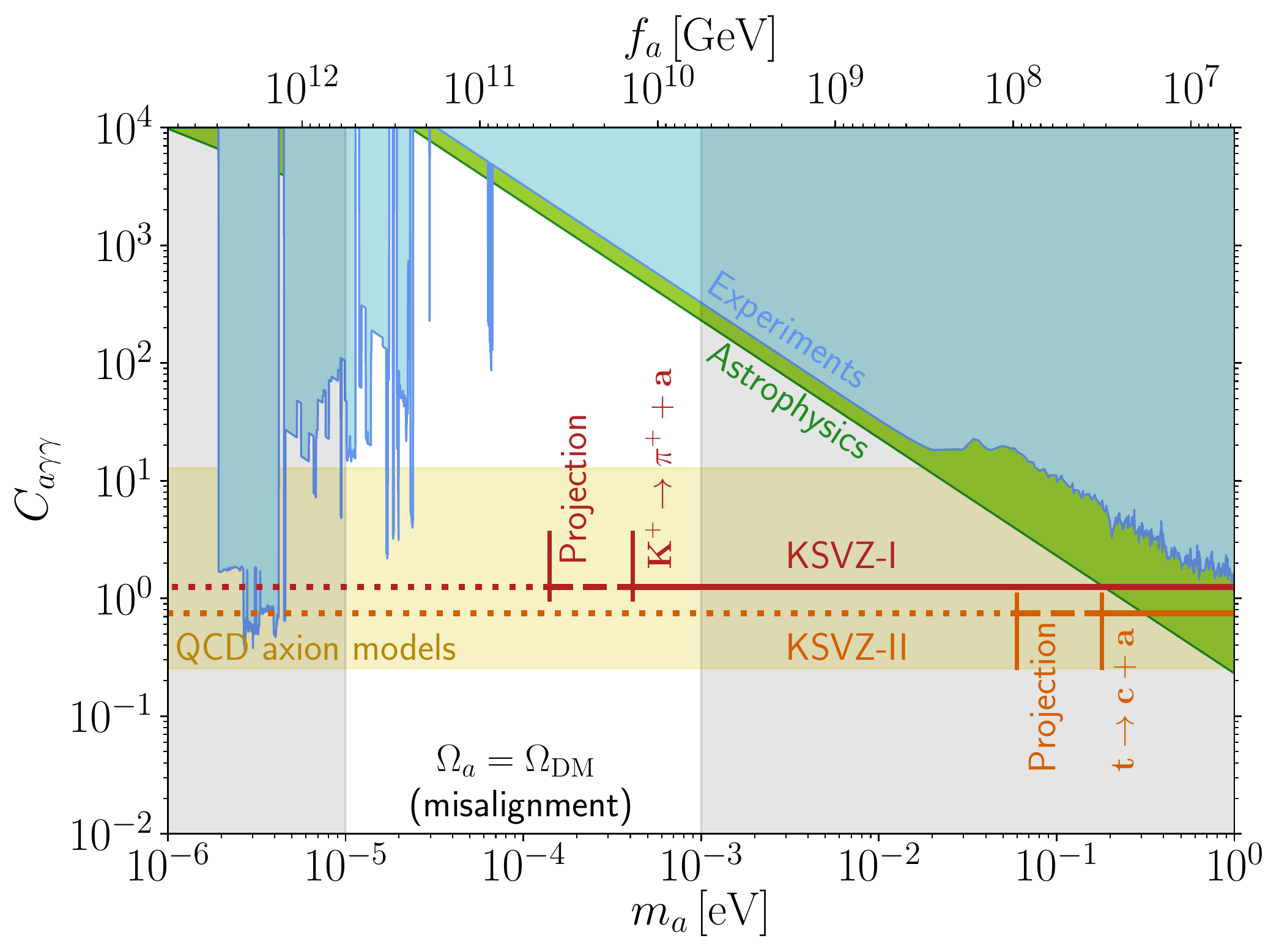}
\caption{Axion mass/decay constant versus photon coupling.
QCD axion models lie in the yellow band, with the two cosmologically viable models shown in red (KSVZ-I) and orange (KSVZ-II); see Eq.\ (\ref{R1R2def}).
The solid and dashed parts of these lines respectively show the current and projected sensitivity of searches for flavor-violating axion couplings.
The region where the QCD axion can make up the observed dark matter abundance is unshaded (see Table~\ref{tab:axion_density}) \cite{Buschmann:2021sdq,Gorghetto:2020qws,Klaer:2017ond,Kawasaki:2014sqa}
Also shown are various other constraints arising from terrestrial experiments~\cite{CAST:2017uph,DePanfilis:1987dk,Hagmann:1990tj,ADMX:2021nhd,Yi:2022fmn,TASEH:2022vvu,Adair:2022rtw,Alesini:2022lnp,Quiskamp:2022pks,HAYSTAC:2023cam,AxionLimits} (blue) and astrophysical observations~\cite{Dolan:2022kul,Noordhuis:2022ljw} (green).
}
\label{fig:fa_ma_vs_gagg}
\end{figure}

To put our results into perspective, Fig.~\ref{fig:fa_ma_vs_gagg} compares this reach with that of experiments looking for dark matter axions through their two-photon interaction with strength $g_{a\gamma\gamma} = C_{a\gamma\gamma}\alpha_{\rm em}/(2\pi f_a)$.
The yellow band in this figure shows the range of photon couplings populated by KSVZ and DFSZ models~\cite{DiLuzio:2016sbl,DiLuzio:2017pfr}, and the red and orange lines show the values of $C_{a\gamma\gamma}$ for the two dark-matter-compatible constructions KSVZ-I and KSVZ-II.
The flavor-violating axion couplings to quarks that arise in these two scenarios are independent of the axion-photon coupling, and thus the rare-decay limits apply directly to $f_a$ (or $m_a$).
The corresponding reach of existing and projected searches is represented by different line styles for the most sensitive flavor combination in each case, and saturating the limits on $C_a^R$.

Figure~\ref{fig:fa_ma_vs_gagg} highlights that, while no existing dedicated axion experiment or astrophysical observation has reached the KSVZ-I and -II lines in the $10\,\mu\mathrm{eV}\leq m_a \leq 10^3\,\mu\mathrm{eV}$ region, rare kaon decay experiments will soon be sensitive to axion masses as small as $100\,\mu\mathrm{eV}$.
Interestingly, searches for flavor-violating axion couplings can more easily test the higher range of QCD axion dark matter masses, which is the most challenging one for axion haloscope experiments due to the required cavity volumes becoming too small.
Therefore, our study constitutes a way to test the QCD axion dark matter parameter space that is complementary to existing direct detection efforts.

\section{Summary and conclusions}
\label{sect:concl}
Hadronic axions provide an elegant solution to the strong CP problem, while also providing an explanation for the dark matter of the universe.
In striking contrast to other axion models, the KSVZ framework can avoid the domain wall problem of early cosmology. 
An essential ingredient of these constructions is a heavy vectorlike quark $Q$, which would
constitute a dangerous stable relic in its absence of couplings to the light quarks. 
These new interactions are only possible in two concrete realizations of the KSVZ scenario, here denoted as KSVZ-I and KSVZ-II, which unavoidably lead to 
mass mixing between the heavy $Q$ and the SM quarks.
Ultimately, this induces chiral interactions of the axion with the light quarks. 
In this work, we have studied the low-energy phenomenology of these generally flavor-violating interactions.
Our main conclusion is that rare meson decay experiments have sensitivity to dark matter axions in the post-inflationary misalignment scenario, as can be seen in Fig.~\ref{fig:fa_ma_vs_gagg}.

After diagonalizing the quark squared-mass matrix, the axion is found to couple derivatively to the SM quarks with strength $C_{qq'}/f_a$, for potentially off-diagonal interactions with flavors $q$ and $q'$. 
However, the mass mixing also induces dimension-4 flavor-violating couplings of the Higgs boson to $q q'$.
The interaction strength is proportional to $C_{qq'}$, allowing us to constrain this parameter independently of $f_a$.
The strongest constraints arise from rare decays such as $t\to h c,\,hu$ and
through its effect on meson-antimeson oscillations.
Figure~\ref{fig:Ca_fa} shows the corresponding limits and their interplay with the ones involving the axion couplings.

Out of all the possible flavor variations, the coupling to the $sd$ combination is found to be most promising.
Experiments like NA62 and KOTO are projected to reach an exquisite precision in the branching ratio for the $K\rightarrow\pi+a$ rare decay (in its charged and neutral versions, respectively).
As an example, NA62 will be sensitive to KSVZ-I QCD axions with masses as small as $100\,\mu\mathrm{eV}$.
This is well inside the $10-10^3\,\mu\mathrm{eV}$region that state-of-the-art calculations indentify as preferred for axions that could make up the dark matter of the universe in the post-inflationary misalignment mechanism.
Excitingly, the NA62 successor at CERN is expected to improve the sensitivity to the $K\rightarrow\pi+a$ branching ratio by at least another factor of two~\cite{Goudzovski:2022vbt}, and thus probe even lighter QCD axions.

\vspace{0.2pt}
\textbf{Acknowledgements.}
We thank  Jacky Kumar and David London for helpful correspondence.
This work was supported by the Natural Sciences and Engineering 
Research Council (NSERC) of Canada.
\vspace{0.2pt}
\section{Appendices}

\subsection{Quark mixing angles}
\label{sec:quark_mixing_angles}
Here we give details of the computation of the mixing angles between light and heavy quarks,
as well as the light ones amongst themselves.
Let $M$ be the mass matrix in Eq.\ (\ref{mixmat}), in block form,
\be
    M = \left({m_d\atop Y^T}{0\atop m_Q}\right)\,,
\ee
where $Y_i = y_{2,i} f_a$ and $m_d$ can be considered as diagonal.
For the right-handed mixing angles, we are interested in the matrix
\be
    M^\dagger M = \left({\tilde m_d^2 \atop m_Q Y^T}{m_Q Y^*\atop m_Q^2}\right)\,,
\ee
where $\tilde m_d^2 = m_d^2 + Y^* Y^T$.
This can be block-diagonalized via $U^\dagger M^\dagger M U$ with the unitary transformation (up to terms of $O(\epsilon^2)$)
\be
    U \cong \left({1- \sfrac12\epsilon\epsilon^\dagger\atop -\epsilon^\dagger}
    {\epsilon\atop 1 - \sfrac12\epsilon^\dagger\epsilon}\right)\,,
\ee
where 
\be
    \epsilon = m_Q(m_Q^2 - \tilde m_d^2 )^{-1} Y^*\,.
\ee

Notice that at this step, the term $-\epsilon\epsilon^\dagger/2$ already gives a contribution to the light quark mixing.  There is a further contribution from completing the diagonalization in the upper $3\times 3$ block.
One finds that
\be
    U^\dagger M^\dagger M U \cong \left({A\atop 0}{0\atop m_Q^2}\right)\,,
\ee
where the upper $3\times 3$ block is
\be
    A = m_d^2 - {1\over 2 m_Q^2}\left\{ Y^* Y^T,\, m_d^2\right\}
\ee
plus terms higher order in $\epsilon$ and $m_d^2/m_Q^2$.  $A$ can be perturbatively diagonalized to find the extra contribution to the right-handed mixing angles.  The total mixing angle takes the form 
\bea
    \theta^R_{d d'} &\cong& -\frac12\left(1 + {m^2_d+m^2_{d'}\over m_d^2 - m_{d'}^2}\right) {Y^*_d Y_{d'}\over m_Q^2}\nn\\
        &=& {m^2_d\over m^2_d-m^2_{d'}}{Y^*_d Y_{d'}\over m_Q^2}\,,
\eea
to leading order in $\epsilon$.

For the left-handed light quark mixing, we diagonalize $M M^\dagger$.
The partial diagonalization to remove the mixing with $Q$ generates a subdominant contribution to the light quark mixing, and an addition to the light quark mass matrix of the form
\be
    m^2_d \to m^2_d - {m_i m_j Y^*_i Y_j\over(m_Q^2 + Y^T Y^*)  }\,,
\ee
where $m_d$ is diagonal.  Perturbatively diagonalizing the extra contribution leads to Eq.\ (\ref{thetaL}).

In the foregoing derivation of the mixing angles, the step of integrating out the heavy $Q$ introduces a loss of unitarity in the
CKM matrix,
\bea
    V^\dagger V &\to& \left(1- \sfrac12\epsilon\epsilon^\dagger\right)
        V^\dagger V  \left(1- \sfrac12\epsilon\epsilon^\dagger\right)
        \nn\\ &=& 1- \epsilon\epsilon^\dagger
\eea
where 
\be
    \epsilon_i = {m_i Y^*_i\over m_Q^2 + Y^T Y^*}
\ee
for the left-handed quark mixing.

\bibliography{bibliography}

\end{document}